# Lightweight, error-tolerant edge detection using memristor-enabled stochastic logics

**Authors**
Lekai Song[1], Pengyu Liu[1], Jingfang Pei[1], Yang Liu[1,2], Songwei Liu[1], Shengbo Wang[3], Leonard W. T. Ng[4], Tawfique Hasan[5], Kong-Pang Pun[1], Shuo Gao[3], Guohua Hu[1,*]

**Affiliations**
[1]Department of Electronic Engineering, The Chinese University of Hong Kong, Shatin, N. T., Hong Kong S. A. R., China
[2]Shun Hing Institute of Advanced Engineering, The Chinese University of Hong Kong, Shatin, N. T., Hong Kong S. A. R., China
[3]School of Instrumentation and Optoelectronic Engineering, Beihang University, Beijing 100191, China
[4]School of Materials Science and Engineering, Nanyang Technological University, Singapore 639798, Singapore
[5]Cambridge Graphene Centre, University of Cambridge, Cambridge CB3 0FA, UK

*Correspondence to: ghhu@ee.cuhk.edu.hk

**Abstract**
The demand for efficient edge vision has spurred the interest in developing stochastic computing approaches for performing image processing tasks. Memristors with inherent stochasticity readily introduce probability into the computations and thus enable stochastic image processing computations. Here, we present a stochastic computing approach for edge detection, a fundamental image processing technique, facilitated with memristor-enabled stochastic logics. Specifically, we integrate the memristors with logic circuits and harness the stochasticity from the memristors to realize compact stochastic logics for stochastic number encoding and processing. The stochastic numbers, exhibiting well-regulated probabilities and correlations, can be processed to perform logic operations with statistical probabilities. This can facilitate lightweight stochastic edge detection for edge visual scenarios characterized with high-level noise errors. As a practical demonstration, we implement a hardware stochastic Roberts cross operator using the stochastic logics, and prove its exceptional edge detection performance, remarkably, with 95% less computational cost while withstanding 50% bit-flip errors. The results underscore the great potential of our stochastic edge detection approach in developing lightweight, error-tolerant edge vision hardware and systems for autonomous driving, virtual/augmented reality, medical imaging diagnosis, industrial automation, and beyond.

**Main text**

In edge visual scenarios, extracting image features to enable efficient user-scene interaction and decision-making has been a challenging topic due to the intensive data deluge and constrained hardware resources. In this context, edge detection is widely employed to extract the key visual cues, such as the shallow features of color, contour and texture, for image understanding and initial decision-making (*1*, *2*). However, the conventional edge detection approaches relying on matrix multiplication and gradient computation in binary can still lead to excessive computational cost and latency against edge hardware integration and deployment (*3*). For instance, the deterministic nature of binary computing decides high-precision data representation that can be redundant for the computation (*4*, *5*). Taking multiplication as an example, as shown in Fig. 1a, the function despite its simplicity requires large-scale circuits and multiple logic operations. The scalability of the circuits and operations as well as the resultant latency can increase considerably as the computation throughput increases (*6*). Worse still, the binary data representation with the position-dependent bit weights can make edge detection highly susceptible to errors (*7*). As exampled in Fig. 1a, even a single bit-flip can corrupt both the input and output. Bit-flips are a common error that can be induced by electromagnetic interference (*8*). Consequently, error-tolerance circuits and algorithms are often necessitated to achieve reliable results, resulting in additional cost and latency.

The above challenges posed by binary computing put forward a demand for lightweight, error-tolerant computing paradigms for performing the edge detection and other image processing tasks in edge visual scenarios (*9*). Among the various computing paradigms, stochastic computing emerges as a promising solution (*10*). Unlike binary computing, stochastic computing represents the data as sequences of random 0s and 1s, known as *stochastic numbers*, wherein the bits hold an equal weight and the probability of the 1s in the sequences determines the value of the stochastic numbers (*11*). This stochastic and probabilistic nature of data representation allows for the implementation of lightweight logic circuits and operations. Again, taking multiplication as the example, as shown in Fig. 1b, the function can be achieved with one single AND gate, termed *stochastic multiplier*. Interestingly, the stochastic multiplier can be applied to arbitrary stochastic numbers of varying lengths without scaling up the circuits (*12*). Meanwhile, importantly, stochastic computing due to its stochastic and probabilistic nature is inherently tolerant of errors. As exampled, though the occurrence of a random bit-flip corrupts the input, the output remains unvaried. The impact of the bit-flip error can even get diminished as the length of the stochastic numbers increases (*11*). As such, with its lightweight and error-tolerant characteristics, stochastic computing holds great promise in addressing the aforementioned edge detection challenges. Though promising, adapting edge detection to stochastic computing faces challenges due to the lack of reliable stochastic logics for stochastic number encoding and processing (*11*).

In this work, we develop stochastic logics using filamentary memristors and apply the stochastic logics to facilitate edge detection. Specifically, we realize stochastic number encoders (SNEs) with memristors for stochastic number encoding, and then integrate the SNEs with logic gates to design stochastic logics for stochastic number processing. Harnessing the inherent stochasticity from the memristors, the SNEs can encode data into stochastic numbers with well-regulated probabilities and correlations, enabling the stochastic logics to perform bitwise logic operations with statistical probabilities in performing stochastic computations for edge detection tasks. To demonstrate how the stochastic logics facilitate edge detection, we implement a hardware stochastic Roberts cross operator and prove its exceptional performance in contour and texture extractions. Remarkably, the demonstration achieves 95% less computational cost while withstanding 50% bit-flip error, highlighting the lightweight and error-tolerant edge detection capability of our approach.

**Stochastic number encoders**

Figure 2a presents the circuit design of SNEs we propose, where each SNE consists of a memristor and a set of comparators. Memristors, as one type of emergent neuromorphic electronic device, often exhibit stochasticity in switching originating from their underlying switching mechanisms (*13*). For example, the switching of filamentary memristors, i.e. the formation and rupture of conduction filaments, relies on the stochastic diffusion of the conduction elements, leading to stochasticity inherent in switching (*14*). Though stochasticity is non-ideal for high-precision computing, we propose it can be harnessed as a reliable entropy source for SNE realization. By integrating a memristor, the SNE in our design can encode the input data into stochastic numbers – the memristor when fed with pulsed signal $V_{in}$ can be switched stochastically, and its output carrying the stochasticity can be binarized by the comparators via the reference $V_{ref}$ for stochastic number encoding with probability. As such, the probability of the stochastic numbers is expected to be well-regulated by $V_{in}$ and $V_{ref}$. Notably, as designed, the SNE via convenient circuit reconfiguration can output stochastic numbers in positive and negative correlations, while two or more parallel SNEs can encode uncorrelated stochastic numbers.

To implement the SNEs, we prepare filamentary memristors from solution-processed hexagonal boron nitride (hBN), following our previous report (*15*). Briefly, hBN is produced by liquid-phase exfoliation (Fig. S1) and used to fabricate the memristors in a Pt/Au/hBN/SiO$_x$/Ag configuration (Fig. S2). We show in Fig. 2b and c an array of the memristors and the structure of a typical device. Note that the use of SiO$_x$ not only preserves the stochastic diffusion of the silver ions but also increases the device fabrication yield to 100% (*15*). In a typical switching (Fig. 2d), the memristor switches to a low resistive state at the threshold voltage $V_{th}$ as the silver ions diffuse and form conduction filaments, and spontaneously resets to a high resistive state once the bias drops below the hold voltage $V_{hold}$. Due to the stochastic diffusion of the silver ions, the switching exhibits stochasticity in both $V_{th}$ and $V_{hold}$. To assess the stochasticity, we conduct over 1,000 consecutive sweeping cycles on the memristor. As shown, the current-voltage output establishes a profound cycle-to-cycle stochasticity in the switching (Fig. 2d), and the corresponding distributions of $V_{th}$ and $V_{hold}$ with Gaussian fittings suggest a stable stochasticity (Fig. 2e). To evaluate the stability of the stochasticity, we perform Ornstein-Uhlenbeck process modelling on the measured $V_{th}$ (Fig. S3). As demonstrated, $V_{th}$ renders as a mean-reverting behavior with random fluctuations, consistent with the simulated Ornstein-Uhlenbeck process, i.e. a stochastic process in a dynamical system (*16*). This indicates the stability of the stochasticity in prolonged switching operations, which is a premise for SNE implementation. Notably, indeed, the endurance test for over 120,000 cycles proves a stable yet stochastic switching of the memristor (Fig. S4).

We integrate the memristors into the circuits (Fig. 2a) to develop the SNEs. See Fig. S5 for the hardware realization of the SNEs. As discussed, the output of the stochastic numbers by the SNEs can be regulated by both $V_{in}$ and $V_{ref}$. Here we show in Fig. 2f the probability of uncorrelated stochastic number $P_{uncorrelated}$ with respect to $V_{in}$. Note that the error bar at each $V_{in}$ representing the variability is obtained from 100 repeated samplings, where each sampling consists of 100 consecutive pulsed signal cycles. As $V_{in}$ increases, $P_{uncorrelated}$ is increased as the memristors tend to be switched on. This proves that the stochastic number occurring at a certain time is probabilistically 0 or 1 and $P_{uncorrelated}$ can be determined by $V_{in}$. Notably $P_{uncorrelated}$ follows a sigmoidal fitting $P_{\text{uncorrelated}} = 1/(1 + exp[-38.9(V_{\text{in}} - 1.34)])$, proving that the SNEs can encode data into stochastic numbers with a well-regulated probability, thereby promising for stochastic computing implementation. In turn, the $P_{\text{uncorrelated}}$-$V_{\text{in}}$ relation can be employed as a guidance to practically regulate $P_{\text{uncorrelated}}$ with $V_{\text{in}}$. Similarly, we show in Fig. 2g the probabilities of positively and negatively correlated stochastic numbers $P_{\text{positive}}$ and $P_{\text{negative}}$ with respect to $V_{\text{ref}}$. $P_{\text{positive}}$ ($P_{\text{negative}}$) decreases (increases) as $V_{\text{ref}}$ increases in positive (negative) correlation, as $V_{\text{ref}}$ serves as the threshold for binarization. Again, $P_{\text{negative}}$ follows a sigmoidal fitting $P_{\text{negative}} = 1/(1 + exp[-63.1(V_{\text{ref}} - 0.19)])$, and $P_{\text{positive}} = 1 - P_{\text{negative}}$. See Fig. S6 for an experimental example of the positively correlated stochastic number encoding. Therefore, the memristor-enabled SNEs prove data encoding into stochastic numbers with regulated probabilities and correlations, facilitating subsequent stochastic logic development.

**Stochastic logics**

We integrate the SNEs with the conventional logic gates to build stochastic logics. To begin with, we investigate the AND and OR gates that are widely used in binary computing. Using stochastic AND logic in uncorrelation as an example, we connect two parallel SNEs to a typical AND gate (Fig. 3a). In this design, the uncorrelated stochastic outputs encoded by the SNEs serve as the inputs to the AND gate, enabling stochastic multiplication of the stochastic outputs. When in operation, based on the demonstrated $P_{\text{uncorrelated}}$-$V_{\text{in}}$ relation in Fig. 2f, the SNEs are fed with pulsed signal cycles of the corresponding $V_{\text{in}}$ to encode uncorrelated stochastic numbers, denoted as $a$ and $b$, with probabilities of $P(a)$ and $P(b)$, respectively. Then, $a$ and $b$ are bit-by-bit fed into the AND gate, yielding a stochastic number output, denoted as $c$, with a probability of $P(c)$. We show in Fig. 3a the corresponding stochastic numbers and probabilities from the experimental hardware test. The statistical relation between the probabilities, i.e., $P(a)P(b) \approx P(c)$, proves that the stochastic AND logic functions as a stochastic multiplier for one-step multiplication of stochastic numbers. Importantly, compared to the binary multiplier in Fig. 1a, this stochastic multiplier significantly simplifies circuit design and reduces computational cost. Besides, the SNEs can be configured to exhibit positive (negative) correlation, enabling positively (negatively) correlated stochastic AND logic operations (Fig. 3a). The output probability $P(c)$ in the correlated cases is determined by the minimum (maximum) value of $P(a)$ and $P(b)$ instead. Table 1 summarizes the statistical relations between $P(a)$, $P(b)$ and $P(c)$ for the stochastic AND logic in all three correlations, proving stochastic logic operations with well-regulated probabilities and correlations. Similarly, we build stochastic OR logic in all three correlations, and it performs different logic operations as designed (Fig. 3b). The statistical relations between $P(a)$, $P(b)$ and $P(c)$ are also summarized in Table 1.

As discussed, edge detection involves matrix multiplication and gradient computation that normally require large-scale binary logics and considerable logic operations (*3*). In contrast, it is possible to perform absolute-valued subtraction for gradient computation with minimal computational and hardware costs using the stochastic logics. Here we propose in Fig. 3c the design of stochastic XOR logic, consisting of only an SNE and an XOR gate, to perform the function. Specifically, the SNE is fed with pulsed signal cycles of the corresponding $V_{\text{in}}$ according to the $P_{\text{positive}}$-$V_{\text{in}}$ relation in Fig. 2g to encode positively correlated stochastic numbers, denoted as $a$ and $b$, with respective probabilities of $P(a)$ and $P(b)$. Then, $a$ and $b$ serve as the inputs to the XOR gate, and the resultant $P(c)$ satisfies $P(c) \approx |P(a) - P(b)|$. In this case, positively correlated stochastic numbers mean a maximum overlap of 0s and 1s, such that the probability for

two 1s or two 0s is $min(P(a), P(b))$ or $min(1 - P(a), 1 - P(b))$. Assume $P(a) > P(b)$, the stochastic XOR logic output $P(c) = 1 - P(b) - (1 - P(a)) = P(a) - P(b)$, and vice versa. This proves the capability of the stochastic XOR logic to perform the absolute-valued subtraction function in only one step. Besides gradient computation, denoising, smoothing, and down-sampling are also essential matrix operations in edge detection. A general approach in performing these functions is to use mean convolutional filters to process the pixels. Here we propose in Fig. 3d (see also Fig. S7) the design of stochastic MUX logic to realize a mean convolutional filter. See the statistical relations for the XOR and MUX logics in Table 1.

The above stochastic logics lay the foundation for performing lightweight and error-tolerant edge detection tasks. Note that in the above demonstrations, the stochastic numbers are encoded in 100-bit for illustrative purposes (see Methods). However, the bit length can be adjusted to accommodate the practical edge detection tasks with different computational precision requirements, given the tradeoff between the computational cost and precision.

**Stochastic edge detection**
As discussed, edge detection in the conventional binary computing approaches relies on the use of large-scale binary logic filters, such as *Roberts cross* and *Sobel* operators, leading to significant computational and hardware costs as well as latency (*3*). In this context, we propose a hardware stochastic Roberts cross operator using the stochastic logics to address the edge detection challenges. Briefly, two SNEs, two XOR gates, and one MUX are integrated to build such an operator. See Fig. 4 and Fig. S8 for the design and hardware realization of the operator.

We apply the stochastic Roberts cross operator in image processing to demonstrate the feasibility of stochastic edge detection. The image for illustrative purposes is captured from the artwork *The Horse in Motion* (Fig. 4a). Here each pixel in 0-255 grayscale is encoded in 100-bits. As shown in Fig. 4b, the stochastic Roberts cross operator is used to scan over the pixel map to yield a gradient map $(i, j)$ reconstructed from the output stochastic numbers. Specifically, one SNE and one XOR gate work consecutively to yield the $x$ component of the output gradient $(i, j)$, denoted as $|Gx|$, while the other SNE and XOR gate yield the $y$ component, denoted as $|Gy|$. The gradient $G(i, j)$ is obtained by averaging $|Gx|$ and $|Gy|$ using the MUX logic, i.e. $G(i, j) = 0.5(|Gx| + |Gy|)$. The coefficient 0.5 scales the gradient within the original grayscale. See Methods for the details. As such, as demonstrated in Fig. 4c, scanning with stochastic Roberts cross operator over the marked image region of 5×5 pixels in Fig. 4a yields a 4×4 pixeled gradient map that evidently demonstrates successful edge detection, as outlined by the red dashed lines. This confirms the feasibility of the stochastic Roberts cross operator in performing edge detection.

As discussed, the bit length of the stochastic numbers can govern the computational precision. To investigate the impact of the bit length on the stochastic Roberts cross operator for edge detection, we encode the pixels of the image frame in Fig. 4a in 4, 16, 64, and 256-bits, respectively. The edge detection results (Fig. 4d) prove that the edges are successfully detected and recognized in all cases. However, as observed, a longer bit length yields better edge detection. To quantitatively evaluate the performance, we compare the edge detection results with those obtained from the standard algorithmic method. We consider the algorithmic result as the ground truth, and assess the fidelity of the stochastic edge detection using two metrics: the structural similarity index measure (SSIM) and peak signal-to-noise ratio (PSNR). Here we visualize the loss in performance by the SSIM maps (Fig. 4e), where a brighter pixel indicates a higher similarity to the ground truth, i.e. a better edge detection performance. This thus reveals and confirms that a longer bit length indeed leads to an improved edge detection performance. For instance, the 256-bit achieves a near-ideal performance, with SSIM >0.95 and PSNR >30 dB. In contrast, the 4-bit exhibits relatively poor performance, as the limited precision in the 4-bit length fails to accurately encode the 0-255 grayscale. However, as evident in Fig. 4 d and e, the 4-bit still successfully detects the edges. As discussed, stochastic computing with short bit lengths can significantly reduce the computational workload, in addition to the minimized hardware cost, in comparison to its binary counterpart. It is estimated that the stochastic edge detection using the 4-bit length may consume only ~5.6% of the computational resources required by binary computing (*17*). Nevertheless, the investigation demonstrates the capability of the stochastic Roberts cross operator to perform edge detection with minimal computational and hardware costs.

We further investigate the error-tolerance capacity of the stochastic cross operator against bit-flips. Specifically, as illustrated in Fig. 5a, a certain level of bit-flips ranging from 0% to 50% is injected into the stochastic numbers (in 256-bit encoding). Again, we adopt the SSIM and PSNR metrics to evaluate the edge detection performance. As evident in Fig. 5b, d and e, the stochastic Roberts cross operator demonstrates successful edge detection in all levels of bit-flip injections. Notably, the operator even retains an SSIM of >0.95 and a PSNR of >30 dB at a 50% bit-flip injection. In comparison, the performance from the standard algorithmic method substantially degrades at a bit-flip injection of only 5%, with the edges hardly recognized and the SSIM and PSNR significantly decreased (Fig. 5c, d and e). See Fig. S9 for the SSIM maps from the standard algorithmic method, and the error-tolerance results at more bit-flip injection levels. The superior error-tolerance capacity of the stochastic Roberts cross operator originates from the fact that each bit in the stochastic numbers carries an equal weight, and thus the impact of pairs of bit-flips can be cancelled.

The stochastic Roberts cross operator, with the exceptional edge detection performance, shows great potential for performing real-world edge detection tasks. To explore the feasibility, we prove its effectiveness in processing the video clip of *The Horse in Motion* and extracting the contours. In addition, we apply the operator to process real-world traffic video, and demonstrate contour extraction from the high-speed vehicles. This opens up the possibility of the operator to enable real-time decision-making in transport planning and driving automation. Besides, we employ the operator to process real-time magnetic resonance imaging characterized by low resolution (*18*). The successful highlight of the vocal tract contours during human speech production indicates the potential for developing lightweight edge medical tools for speech and language rehabilitation.

**Conclusion**

In this work, we have presented a lightweight, error-tolerant stochastic edge detection approach using memristor-enabled stochastic logics. The stochastic logics, realized by integrating solution-processed hBN filamentary memristors into compact logic circuits, allow for stochastic number encoding and processing with well-regulated probabilities and correlations. Owing to these characteristics, the stochastic logics can facilitate lightweight and error-tolerant stochastic edge detection for edge visual scenarios characterized with noise errors. As a practical demonstration, we have designed and implemented a hardware stochastic Roberts cross operator using the stochastic logics, and proved its exceptional edge detection performance, with excellent error-tolerance capacity and low computational cost. Given the remarkable edge detection performance, and the scalability of the memristors and the compacity of the circuit designs, our stochastic logics hold the potential to be readily integrated and generalized towards the development of lightweight and error-tolerant edge vision hardware and systems for edge visual applications from autonomous driving and virtual/augmented reality to industrial automation and medical imaging diagnosis.

## Methods

**Filamentary memristors:** Pristine hBN powder and all chemicals are purchased from Sigma-Aldrich and used as received. Liquid-phase exfoliation, ink formulation, and deposition of hBN follow our method previously reported in Ref. (*15*). In a typical process, the memristor is fabricated in a vertical Pt/Au/hBN/SiO$_x$/Ag configuration, where hBN is deposited by slot-die coating, the SiO$_x$ layer (10 nm) is deposited by electron beam evaporation, and the metal electrodes (5/15 nm Pt/Au and 30 nm Ag) are patterned by photolithography and deposited by electron beam evaporation. The device substrate is Si/SiO$_2$. The slot-die coater is Ossila L2005A1. The evaporator is IVS EB-600. During device fabrication, the hBN layer after deposition is baked at 200ºC for 2 hours.

**SNEs and stochastic logics:** To build the SNEs and stochastic logics, the memristors are tested on a probe station and connected to the logic gates and other electronic devices on a breadboard. See the hardware realization in Fig. S5 and S8. Tektronix Keithley 4200A-SCS parameter analyzer with pulse measure units is used to measure the electrical characteristics of the memristor. Siglent arbitrary waveform generators and digital storage oscilloscope are used to output the signals and measure the output waveforms. To endow the stochastic numbers with a certain probability, based on the demonstrated $P_{\text{positive}}$-$V_{\text{in}}$ relation in Fig. 2g, each SNE are fed with $n$ pulsed signal cycles of the corresponding $V_{\text{in}}$ to encode $n$-bit stochastic numbers. The bit length is determined by $n$.

**Edge detection:** The Roberts cross operator consists of two 2×2 kernels, i.e. $\begin{bmatrix} 1 & 0 \\ 0 & -1 \end{bmatrix}$ and $\begin{bmatrix} 0 & 1 \\ -1 & 0 \end{bmatrix}$. As stochastic computing works on the probability domain, each pixel $(i, j)$ of a grayscale image needs to be initially normalized into a probability, denoted as $P(i, j)$. Hence, for a localized pixel map $\begin{bmatrix} P(i, j) & P(i + 1, j) \\ P(i, j + 1) & P(i + 1, j + 1) \end{bmatrix}$, each SNE in the stochastic Roberts cross operator is used to encode two positively correlated stochastic numbers, the probabilities of which correspond to the diagonal values in the localized pixel map. As such, two SNEs of the operator encode two pairs of positively correlated stochastic numbers that serve as the inputs to the XOR gates. A pair is input into the XOR gate to yield the $x$ of the output gradient at pixel $(i, j)$, denoted as $|Gx| = |P(i, j) - P(i + 1, j + 1)|$, while the other pair and XOR gate yield the $y$ component, denoted as $|Gy| = |P(i + 1, j) - P(i, j + 1)|$. $|Gx|$ and $|Gy|$ are then averaged by the MUX to obtain the absolute magnitude of the approximate gradient. See the working principle of stochastic MUX logic in Fig. S7. Limited by the scalability of the lab-based realization of the stochastic Roberts cross operator, large-scale edge detection on images and videos is conducted via simulation in Python3.

## References

1.    R. Shad, J. P. Cunningham, E. A. Ashley, C. P. Langlotz, W. Hiesinger, Designing clinically translatable artificial intelligence systems for high-dimensional medical imaging. *Nat. Mach. Intell.* **3**, 929–935 (2021).

2. B. Peters, N. Kriegeskorte, Capturing the objects of vision with neural networks. *Nat. Hum. Behav.* **5**, 1127–1144 (2021).

3. J. Canny, A computational approach to edge detection. *IEEE Trans. Pattern Anal. Mach. Intell.* **PAMI-8**, 679–698 (1986).

4. D. Kudithipudi, A. Daram, A. M. Zyarah, F. T. Zohora, J. B. Aimone, A. Yanguas-Gil, N. Soures, E. Neftci, M. Mattina, V. Lomonaco, C. D. Thiem, B. Epstein, Design principles for lifelong learning AI accelerators. *Nat. Electron.* **6**, 807–822 (2023).

5. F. Aguirre, A. Sebastian, M. Le Gallo, W. Song, T. Wang, J. J. Yang, W. Lu, M.-F. Chang, D. Ielmini, Y. Yang, A. Mehonic, A. Kenyon, M. A. Villena, J. B. Roldán, Y. Wu, H.-H. Hsu, N. Raghavan, J. Suñé, E. Miranda, A. Eltawil, G. Setti, K. Smagulova, K. N. Salama, O. Krestinskaya, X. Yan, K.-W. Ang, S. Jain, S. Li, O. Alharbi, S. Pazos, M. Lanza, Hardware implementation of memristor-based artificial neural networks. *Nat. Commun.* **15**, 1974 (2024).

6. S. Harris, D. Harris, *Digital design and computer architecture* (Morgan Kaufmann, 2015).

7. Y. Kim, R. Daly, J. Kim, C. Fallin, J. H. Lee, D. Lee, C. Wilkerson, K. Lai, O. Mutlu, Flipping bits in memory without accessing them. *ACM SIGARCH Comput. Archit. News.* **42**, 361–372 (2014).

8. M. Riahi Alam, M. H. Najafi, N. Taherinejad, M. Imani, L. Peng, Stochastic computing for reliable memristive in-memory computation. *Proc. ACM Gt. Lakes Symp. VLSI, GLSVLSI*, 397–401 (2023).

9. H. Jaeger, B. Noheda, W. G. van der Wiel, Toward a formal theory for computing machines made out of whatever physics offers. *Nat. Commun.* **14**, 4911 (2023).

10. P. Li, D. J. Lilja, W. Qian, K. Bazargan, M. D. Riedel, Computation on stochastic bit streams digital image processing case studies. *IEEE Trans. Very Large Scale Integr. Syst.* **22**, 449–462 (2014).

11. A. Alaghi, W. Qian, J. P. Hayes, The promise and challenge of stochastic computing. *IEEE Trans. Comput. Des. Integr. Circuits Syst.* **37**, 1515–1531 (2018).

12. Z. Lin, G. Xie, S. Wang, J. Han, Y. Zhang, A review of deterministic approaches to stochastic computing. *2021 IEEE/ACM Int. Symp. Nanoscale Archit. NANOARCH 2021*, 1–6 (2021).

13. Q. Xia, J. J. Yang, Memristive crossbar arrays for brain-inspired computing. *Nat. Mater.* **18**, 309–323 (2019).

14. J. Tang, F. Yuan, X. Shen, Z. Wang, M. Rao, Y. He, Y. Sun, X. Li, W. Zhang, Y. Li, B. Gao, H. Qian, G. Bi, S. Song, J. J. Yang, H. Wu, Bridging Biological and Artificial Neural Networks with Emerging Neuromorphic Devices: Fundamentals, Progress, and Challenges. *Adv. Mater.* **31**, e1902761 (2019).

15. L. Song, P. Liu, J. Pei, F. Bai, Y. Liu, S. Liu, Y. Wen, L. W. T. Ng, K. P. Pun, S. Gao, M. Q. H. Meng, T. Hasan, G. Hu, Spiking neurons with neural dynamics implemented using stochastic memristors. *Adv. Electron. Mater.* **2300564**, 1–9 (2023).

16. S. Dutta, G. Detorakis, A. Khanna, B. Grisafe, E. Neftci, S. Datta, Neural sampling machine with stochastic synapse allows brain-like learning and inference. *Nat. Commun.* **13**, 2571 (2022).

17. M. Ranjbar, M. E. Salehi, M. H. Najafi, Using stochastic architectures for edge detection algorithms. *ICEE 2015 - Proc. 23rd Iran. Conf. Electr. Eng.* **10**, 723–728 (2015).

18. Y. Lim, A. Toutios, Y. Bliesener, Y. Tian, S. G. Lingala, C. Vaz, T. Sorensen, M. Oh, S. Harper, W. Chen, Y. Lee, J. Töger, M. L. Monteserin, C. Smith, B. Godinez, L. Goldstein,

D. Byrd, K. S. Nayak, S. S. Narayanan, A multispeaker dataset of raw and reconstructed speech production real-time MRI video and 3D volumetric images. *Sci. Data*. **8**, 1–14 (2021).

**Acknowledgements**

**Funding:** GHH acknowledges support from CUHK (4055115) and RGC (24200521), JFP from RGC (24200521), YL from SHIAE (RNE-p3-21), LWTN from NTU (9069) and MOE AcRF Tier 1 (10658), and SG from National Key Research and Development Program of China (2023YFB3208003).

**Author contributions:** LKS, PYL, GHH designed the experiments. LKS, PYL, JFP, YL, SWL performed the experiments. LKS, PYL, GHH analyzed the data. LKS, GHH prepared the figures. LKS, GHH wrote the manuscript. All authors discussed the results from the experiments and commented on the manuscript.

**Competing interests:** The authors declare no competing financial interests.

**Data and materials availability:** The data that support the findings of this study are available from the corresponding authors upon request.

## Figures and Table

**a  Binary computing**

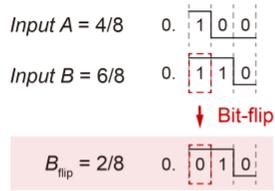

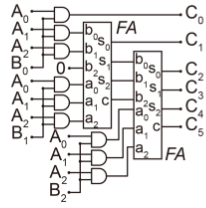

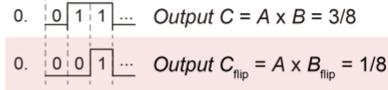

*Binary multiplier*

*Input A = 4/8*    0. 1 0 0

*Input B = 6/8*    0. 1 1 0

↓ Bit-flip

$B_{flip}$ = 2/8    0. 0 1 0

0. 0 1 1 ....    *Output C = A × B = 3/8*

0. 0 0 1 ....    *Output $C_{flip}$ = A × $B_{flip}$ = 1/8*

*FA* is short for Full Adder.

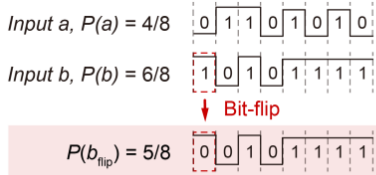

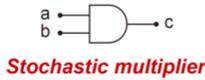

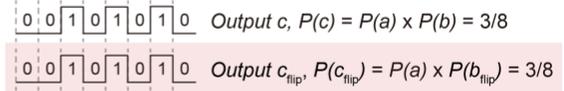

*Input a, P(a) = 4/8*    0 1 1 0 1 0 1 0

*Input b, P(b) = 6/8*    1 0 1 0 1 0 1 1

↓ Bit-flip

$P(b_{flip})$ = 5/8    0 0 1 0 1 1 1 1

*Stochastic multiplier*

0 0 1 0 1 0 1 0    *Output c, P(c) = P(a) × P(b) = 3/8*

0 0 1 0 1 0 1 0    *Output $c_{flip}$, $P(c_{flip})$ = P(a) × $P(b_{flip})$ = 3/8*

**b  Stochastic computing**

**Figure 1. Binary computing vs. stochastic computing.** (a) Binary computing. Two exampled 3-bit binary fraction inputs $A$ and $B$, representing 4/8 and 6/8, respectively, are computed to yield a binary multiplication output $C$ of 3/8. When input $B$ undergoes a bit-flip and the value is changed from 6/8 to 2/8, the output is altered from 3/8 to 1/8. $FA$ is short for *Full Adder*. (b) Stochastic computing. Two exampled 8-bit stochastic number inputs $a$ and $b$, with probabilities $P(a)$ and $P(b)$ of 4/8 and 6/8, respectively, are computed to yield a stochastic multiplication output $P(c)$ of 3/8. When input $b$ undergoes a bit-flip and the value is changed from 6/8 to 5/8, the output remains at 3/8.

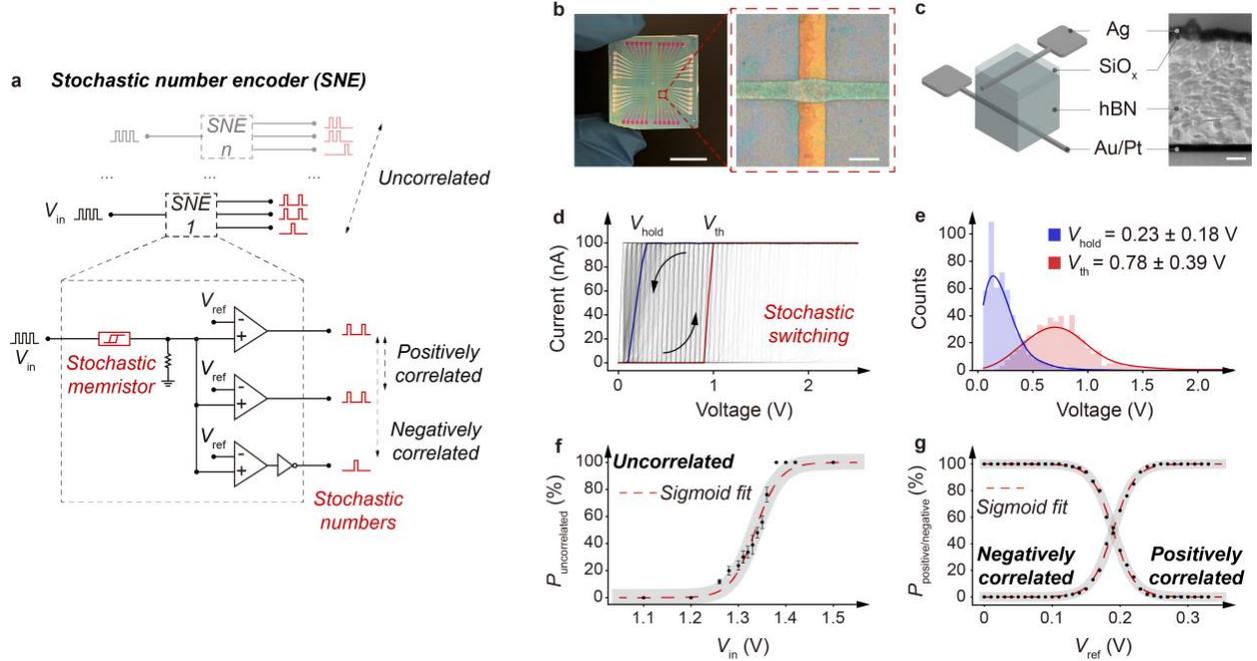

**Figure 2. Stochastic number encoder (SNE).** (a) Schematic SNE, consisting of a memristor and a set of comparators. The output probability and correlation are regulated by both the input $V_{in}$ and reference $V_{ref}$. For negative correlation, a NOT gate is connected to the comparator, and the voltage supply of the NOT gate is synchronized with $V_{in}$ to the memristors to avoid output during the pulse intervals. Independent parallel SNEs are integrated to yield uncorrelated stochastic numbers. See Fig. S5 for the hardware realization of SNE. (b) 12×12 memristor array in a crossbar configuration, with a fabrication yield of 100%. A typical device area is ~20 × 20 μm². Scale bar – 1 cm and 20 μm. (c) Schematic and cross-sectional transmission electron microscopic image of a typical memristor. Scale bar – 50 nm. (d) Current-voltage output from a typical memristor, showing 1,000-cycle stochastic yet stable switching with a ratio of ~$10^5$. $V_{hold}$ and $V_{th}$ denote the hold voltage and threshold voltage. (e) Distributions of the measured $V_{hold}$ ($0.23 \pm 0.18$ V) and $V_{th}$ ($0.78 \pm 0.39$ V), along with the corresponding Gaussian fittings. (f) $P_{uncorrelated}$-$V_{in}$ relation of a typical SNE in uncorrelation, fitting a sigmoid function $P_{uncorrelated} = 1/(1 + exp[-38.9(V_{in} - 1.34)])$. (g) $P_{positive}$-$V_{in}$ and $P_{negative}$-$V_{in}$ relations of the SNE in positive and negative correlations. $P_{negative}$-$V_{in}$ fits a sigmoid function $P_{negative} = 1/(1 + exp[-63.1(V_{in} - 0.19)])$ and $P_{positive} = 1 - P_{negative}$.

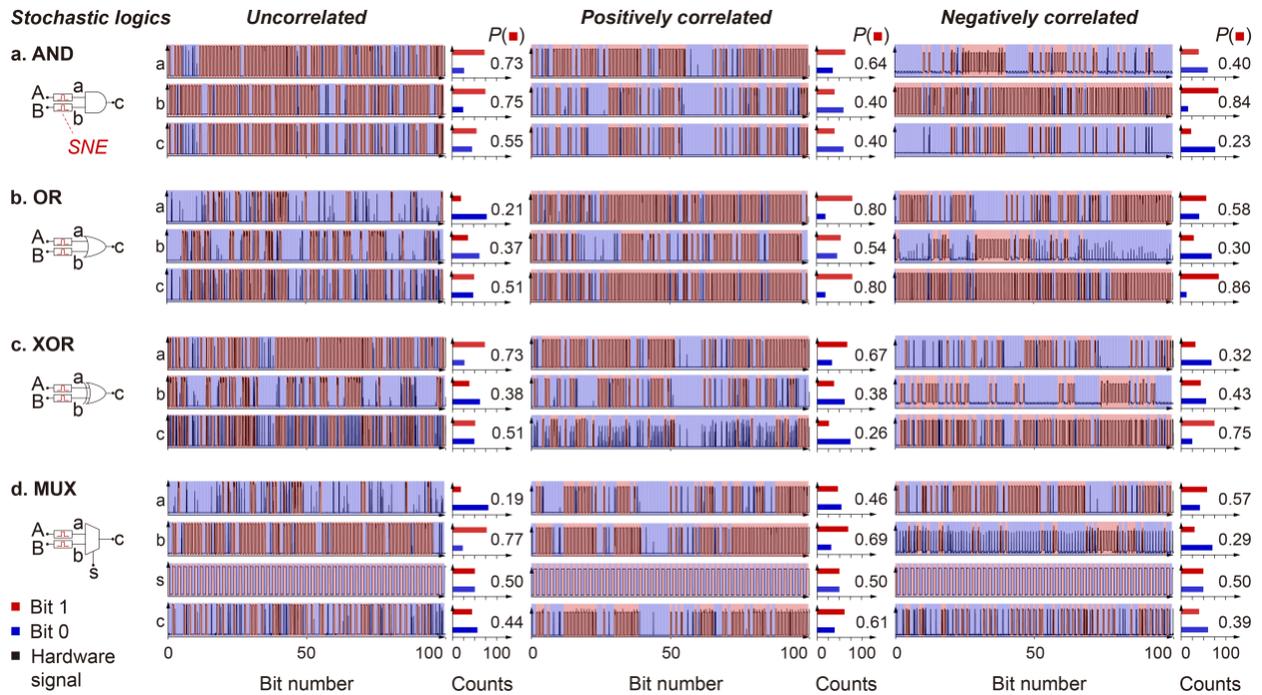

**Figure 3. Stochastic logics.** Schematic stochastic logics in uncorrelation implemented with two independent SNEs and (a) AND, (b) OR, (c) XOR, and (d) MUX, and the corresponding circuit tests of the stochastic logic operations. The stochastic logics can be reconfigured in the positive and negative correlations to yield the stochastic logic operations as respectively demonstrated. For stochastic MUX, the frequency of the select *s* is half of that of the inputs to ensure that both the inputs participate in the logic operations. $P(\blacksquare)$ represents the probability of the 1s in the sequences, i.e. the value of the stochastic numbers. The outputs of stochastic logics in uncorrelation, positive correlation and negative correlation are consistent with the statistical formulas in Table 1.

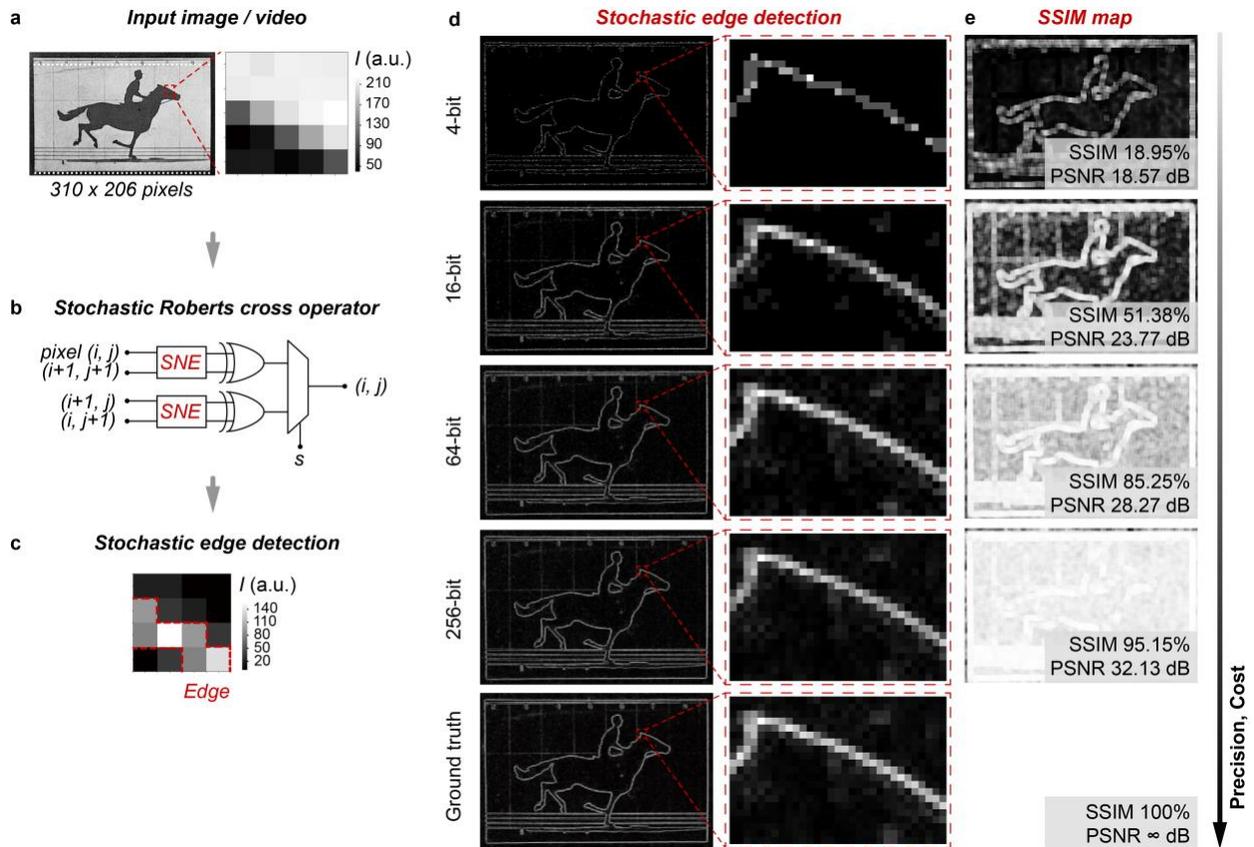

**Figure 4. Stochastic edge detection.** (a) The exampled image, i.e. the first frame of the 'The Horse in Motion', for edge detection demonstration. The region as marked is used to illustrate the edge detection process with the operator. The pixels in 0-255 grayscale are encoded into 100-bits. (b) Schematic stochastic Roberts cross operator, consisting of two SNEs, two XORs, and one MUX. See Fig. S8 for the hardware realization of the operator. (c) Gradient map yielded from scanning with the operator, showing successful edge detection. (d) Edge detection of the first frame with the operator, and (e) the corresponding structural similarity index measure (SSIM) maps and peak signal-to-noise ratios (PSNR). The pixels are encoded into 4, 16, 64, and 256-bits as the inputs. The SSIM and PSNR show that the operator using more bits gives higher edge detection precision. For comparison, the edge detection performed using the standard algorithmic method is presented as the ground truth.

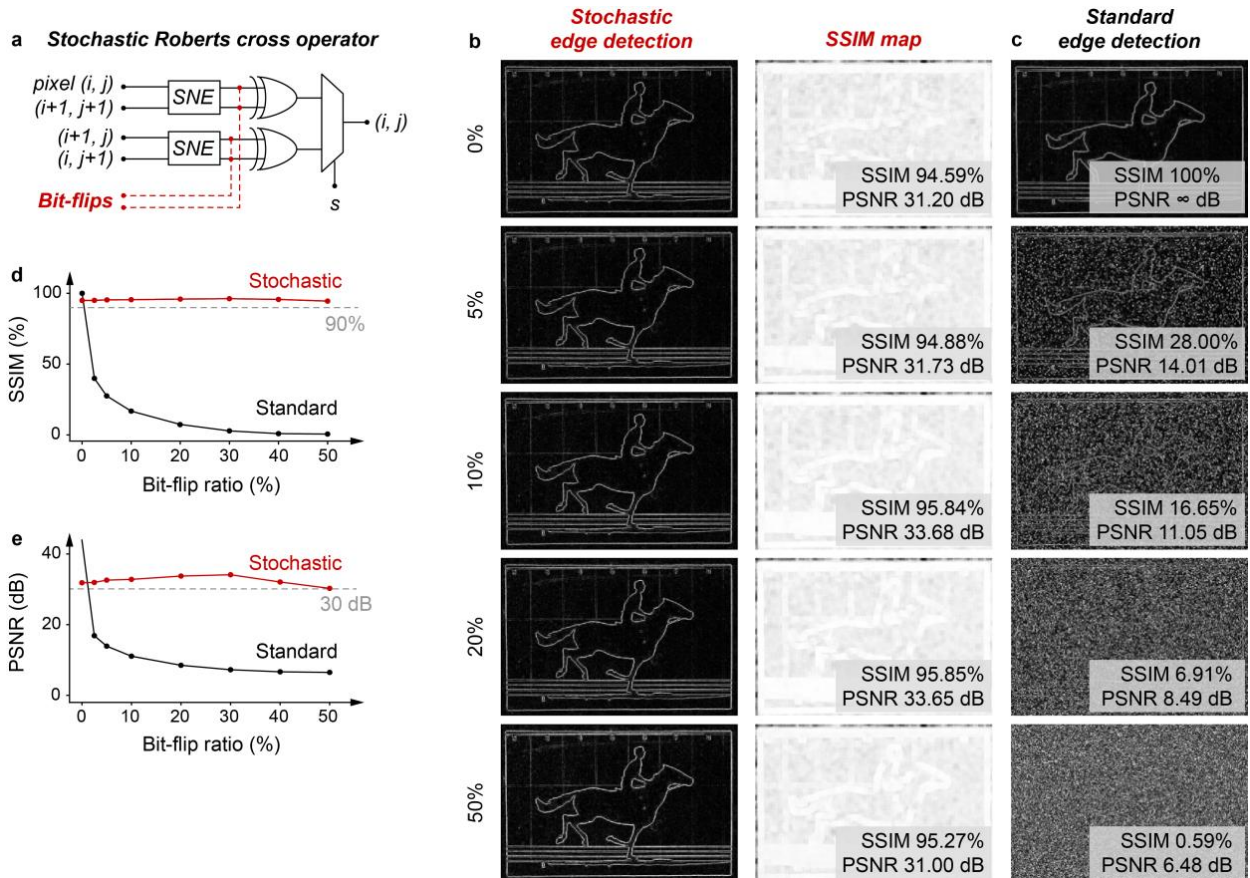

**Figure 5. Error-tolerance test.** (a) Error-tolerance test of the stochastic cross operator, showing bit-flips are injected into the original stochastic numbers for the tests. (b) Stochastic and (c) standard edge detection results and the corresponding SSIM maps of the first frame with bit-flip injection at a ratio of 0%, 5%, 10%, 20%, and 50%. For the stochastic edge detection, the high SSIM (>90%) and PSNR (>30 dB) prove that the bit-flip injection does not degrade the edge detection performance. In contrast, a low level of bit-flip injection significantly degrades the performance of the standard algorithmic edge detection. See Fig. S9 for the SSIM map of the standard edge detection, and the error-tolerance results at more bit-flip injection levels. Performance comparison in the (d) SSIM and (e) PSNR between the stochastic and standard edge detection results.

**Table 1. Statistical formulas of the stochastic logics.** Stochastic logic operations with AND, OR, XOR, and MUX in the uncorrelation, positive correlation, and negative correlation configurations are presented. The stochastic numbers are assumed in a unipolar format (*11*).

| | Uncorrelated | Positively correlated | Negatively correlated |
|---|---|---|---|
| AND | $P(c) = P(a)P(b)$ | $P(c) = \text{Min}(P(a), P(b))$ | $P(c) = \text{Max}(P(a) + P(b) - 1, 0)$ |
| OR | $P(c) = P(a) + P(b) - P(a)P(b)$ | $P(c) = \text{Max}(P(a), P(b))$ | $P(c) = \text{Min}(1, P(a) + P(b))$ |
| XOR | $P(c) = P(a) + P(b) - 2P(a)P(b)$ | $P(c) = \|P(a) - P(b)\|$ | $P(c) = P(a) + P(b)$, if $P(a) + P(b) \leq 1$; $P(c) = 2 - (P(a) + P(b))$, otherwise. |
| MUX | $P(c) = (1 - P(s))P(a) + P(s)P(b)$, if $s$ is uncorrelated with $a$ and $b$ | | |

Supplementary Materials for
**Lightweight, error-tolerant edge detection using memristor-enabled stochastic logics**

**Authors**
Lekai Song[1], Pengyu Liu[1], Jingfang Pei[1], Yang Liu[1,2], Songwei Liu[1], Shengbo Wang[3], Leonard W. T. Ng[4], Tawfique Hasan[5], Kong-Pang Pun[1], Shuo Gao[3], Guohua Hu[1,*]

**Affiliations**
[1]Department of Electronic Engineering, The Chinese University of Hong Kong, Shatin, N. T., Hong Kong S. A. R., China
[2]Shun Hing Institute of Advanced Engineering, The Chinese University of Hong Kong, Shatin, N. T., Hong Kong S. A. R., China
[3]School of Instrumentation and Optoelectronic Engineering, Beihang University, Beijing 100191, China
[4]School of Materials Science and Engineering, Nanyang Technological University, Singapore 639798, Singapore
[5]Cambridge Graphene Centre, University of Cambridge, Cambridge CB3 0FA, UK

*Correspondence to: ghhu@ee.cuhk.edu.hk

**This file contains:**
Supplementary Figure S1-S9
Supplementary References

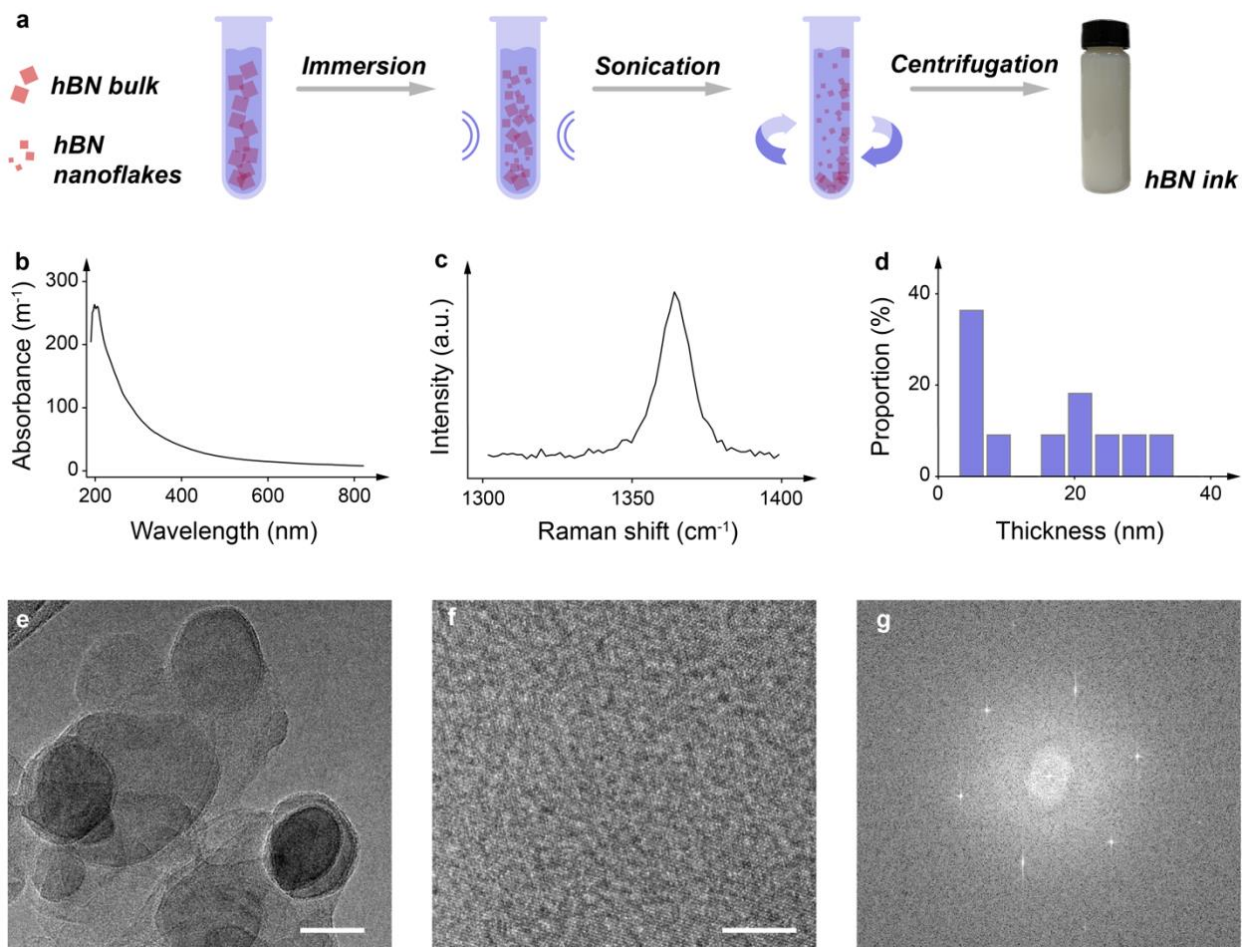

**Figure S1. hBN ink preparation and characterizations.** (a) Schematic liquid-phase exfoliation of hBN, following our previous report Ref. (*1*). After 48-hour bath sonication, the dispersion of the as-exfoliated hBN nanoflakes in isopropanol is centrifuged at 4,000 rpm for 30 minutes. The supernatant is then carefully decanted and collected. The dispersion is then added with controlled volumes of isopropanol and 2-butanol to formulate an ink in isopropanol/2-butanol (90 vol.%/10 vol.%). The process is optimized to give an ink with a concentration of ~1 mg/ml. (b) Absorption spectrum of a typical hBN ink (diluted to 10%), showing no featureless absorption peaks. (c) Raman spectrum of the as-exfoliated exfoliated hBN nanoflakes. (d) Histogram of the thickness of the as-exfoliated hBN nanoflakes characterized by atomic force microscopy (AFM). The averaged thickness is ~9.31 nm, excluding the thick (>20 nm) aggregates formed during the AFM sample preparation process. (e, f) Transmission electron microscopic (TEM) images of the as-exfoliated hBN nanoflakes, and (g) the corresponding fast Fourier transform pattern, confirming the hexagonal lattice structure of as-exfoliated hBN nanoflakes. Scale bars – (e) 50 nm, and (f) 5 nm.

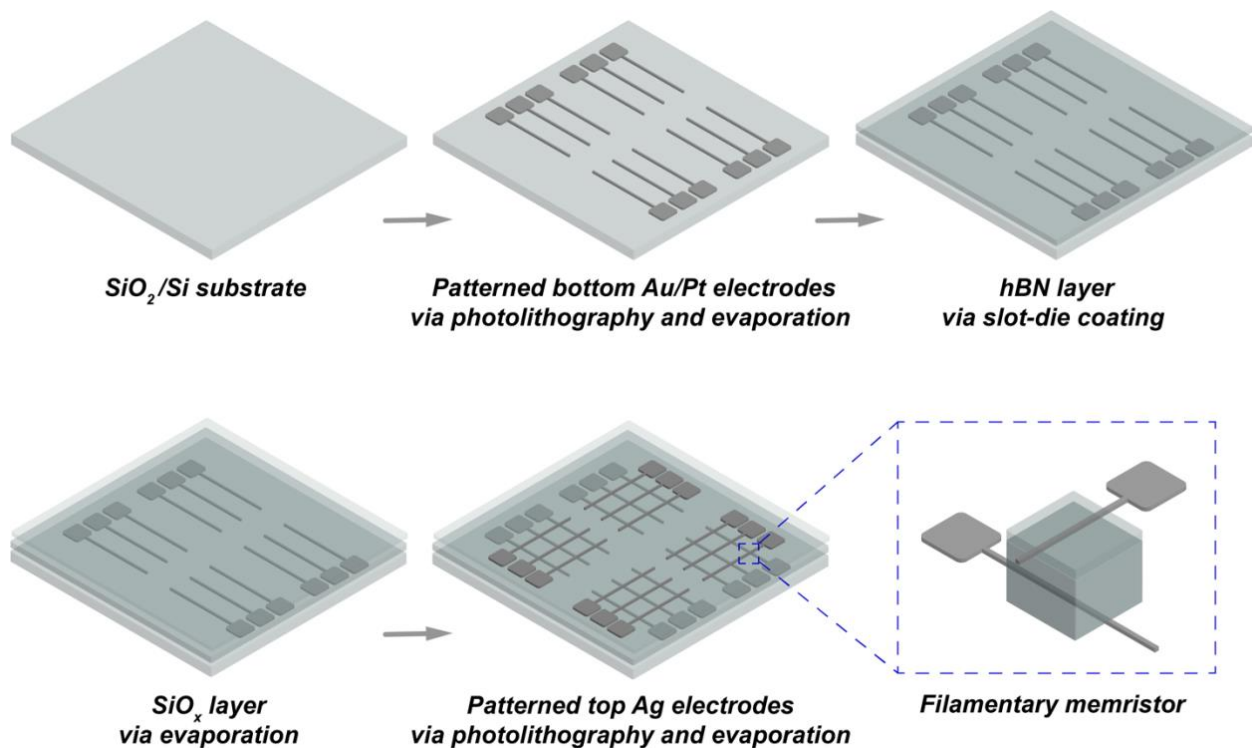

**SiO₂/Si substrate**

**Patterned bottom Au/Pt electrodes
via photolithography and evaporation**

**hBN layer
via slot-die coating**

**SiOₓ layer
via evaporation**

**Patterned top Ag electrodes
via photolithography and evaporation**

**Filamentary memristor**

**Figure S2. Memristor fabrication.** Starting from a cleansed $SiO_2$/Si substrate, the bottom Au/Pt electrodes are patterned using photolithography and evaporation. A hBN layer is deposited on the bottom electrodes using slot-die coating. A $SiO_x$ layer is then deposited using evaporation to minimize the wash-off of hBN during the subsequent patterning processes and increase the fabrication yield to 100%, while preserving the switching behavior of the filamentary memristors. Next, the top silver electrodes are patterned using photolithography and evaporation. As such, the memristor devices are developed at the cross-points of the top and bottom electrodes, in a vertical Au/Pt/hBN/$SiO_x$/Ag configuration.

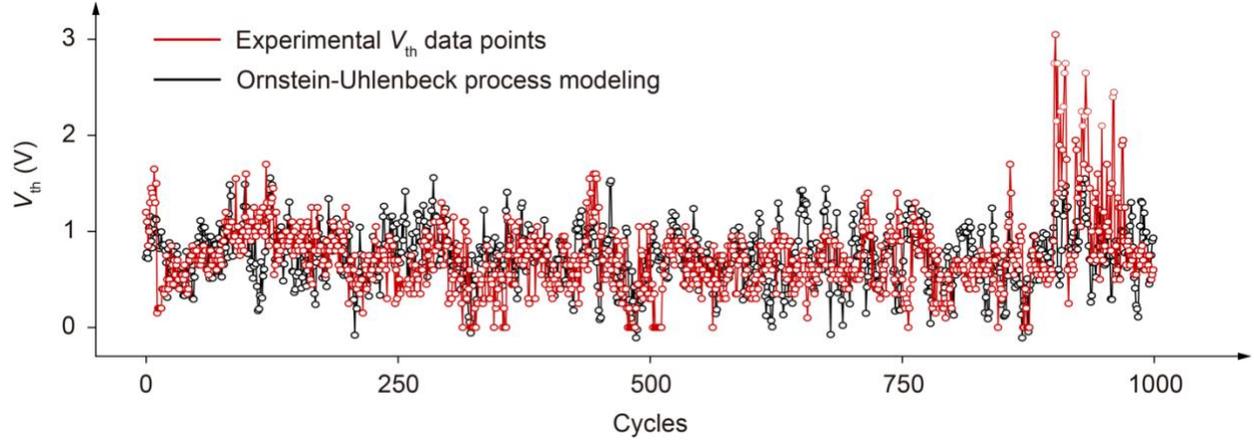

**Figure S3. Stability test of the memristor switching stochasticity.** Ornstein-Uhlenbeck process modeling on the measured threshold voltage $V_{th}$ of the hBN filamentary memristor across the 1,000 consecutive sweeping cycles present in Fig. 2d. The experimental $V_{th}$ data points well fit those from the Ornstein-Uhlenbeck process modelling, where $dV_{th,t} = \theta(\mu - V_{th,t}) + \sigma dW_t = 0.306 \times (0.729 - V_{th,t}) + 0.284 \times dW_t$. An Ornstein-Uhlenbeck process describes a stochastic process in a dynamical system (*2*). A $dW_t$ denotes the variation of a Wiener process, i.e. a real-valued continuous-time stochastic process. This proves the stability of the switching stochasticity of our memristors in prolonged switching operations. At around 900-th cycle, the device is momentarily stuck at the high resistance states and thus hard to switch on probably due to the ambient disturbances but then returns to normal.

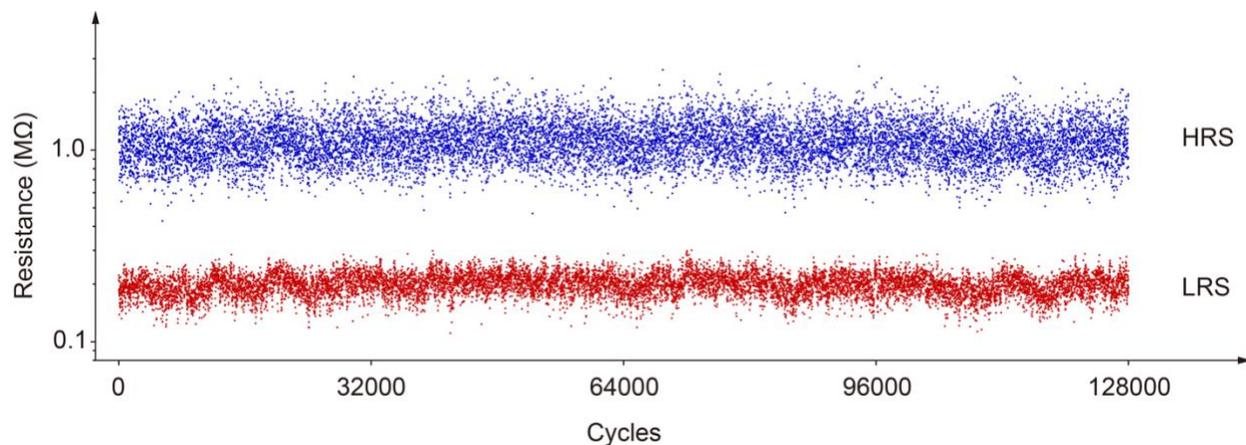

**Figure S4. Endurance test.** Endurance test of a typical hBN filamentary memristor undergoing 128,000 consecutive test cycles. For each test cycle, a 4V/0.2 ms voltage pulse is set to program the memristor and a 0.02V/0.6 ms voltage pulse is set to read the current output. Based on the current output, the high (HRS) and low (LRS) resistance states in each test cycle are measured and plotted as the blue and red dots, respectively. HRS and LRS are clearly distinguished throughout the full test, proving a reliable endurance of our memristors.

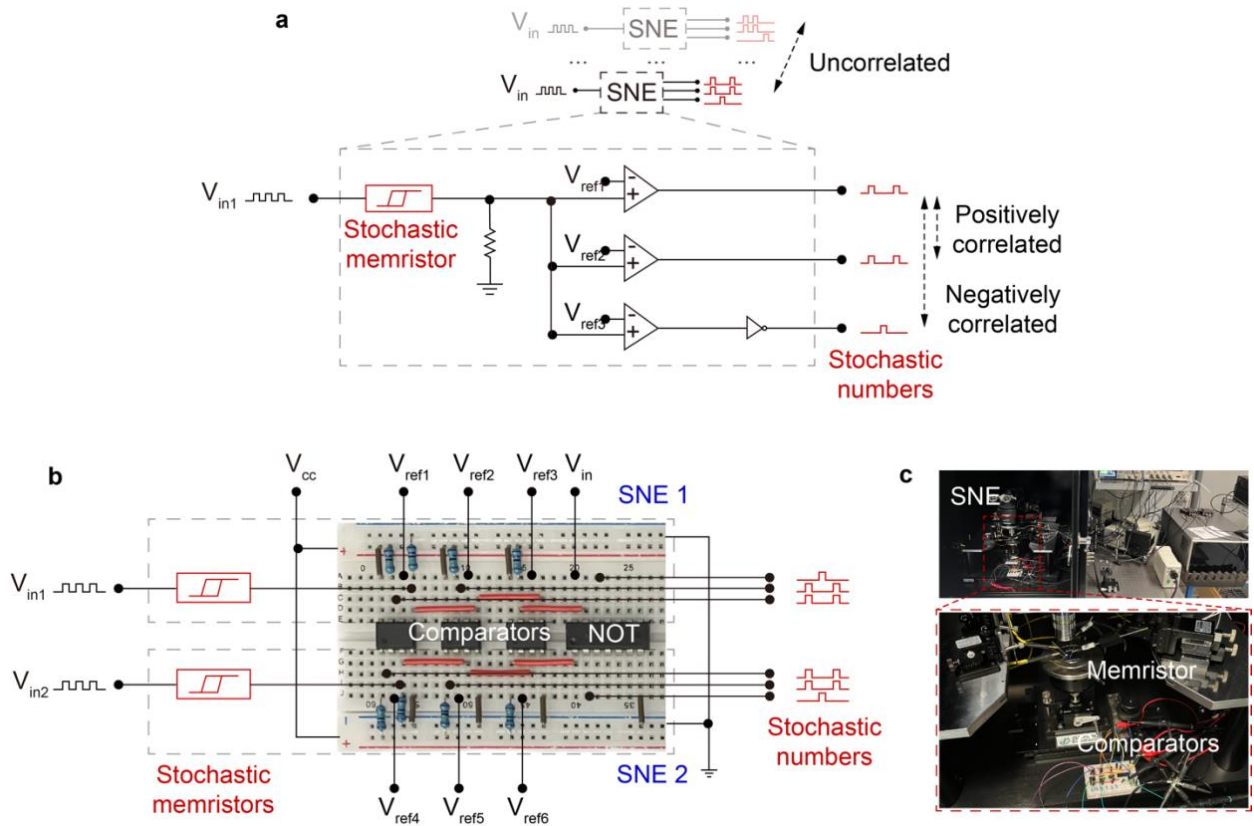

**Figure S5. The circuit design and experimental hardware realization of the SNEs.** (a) Circuit design (replotted from Fig. 2a) and (b) hardware realization of SNEs. (c) Experimental setup. To build the SNEs, the memristors are tested on a probe station and connected to the logic gates and other electronic devices on a breadboard. The electronic components for the SNE realization include comparators, NOT gates, and resistors. Note that the voltage supply of the NOT gates is synchronized with $V_{in}$ to the memristors to avoid output during the pulse intervals.

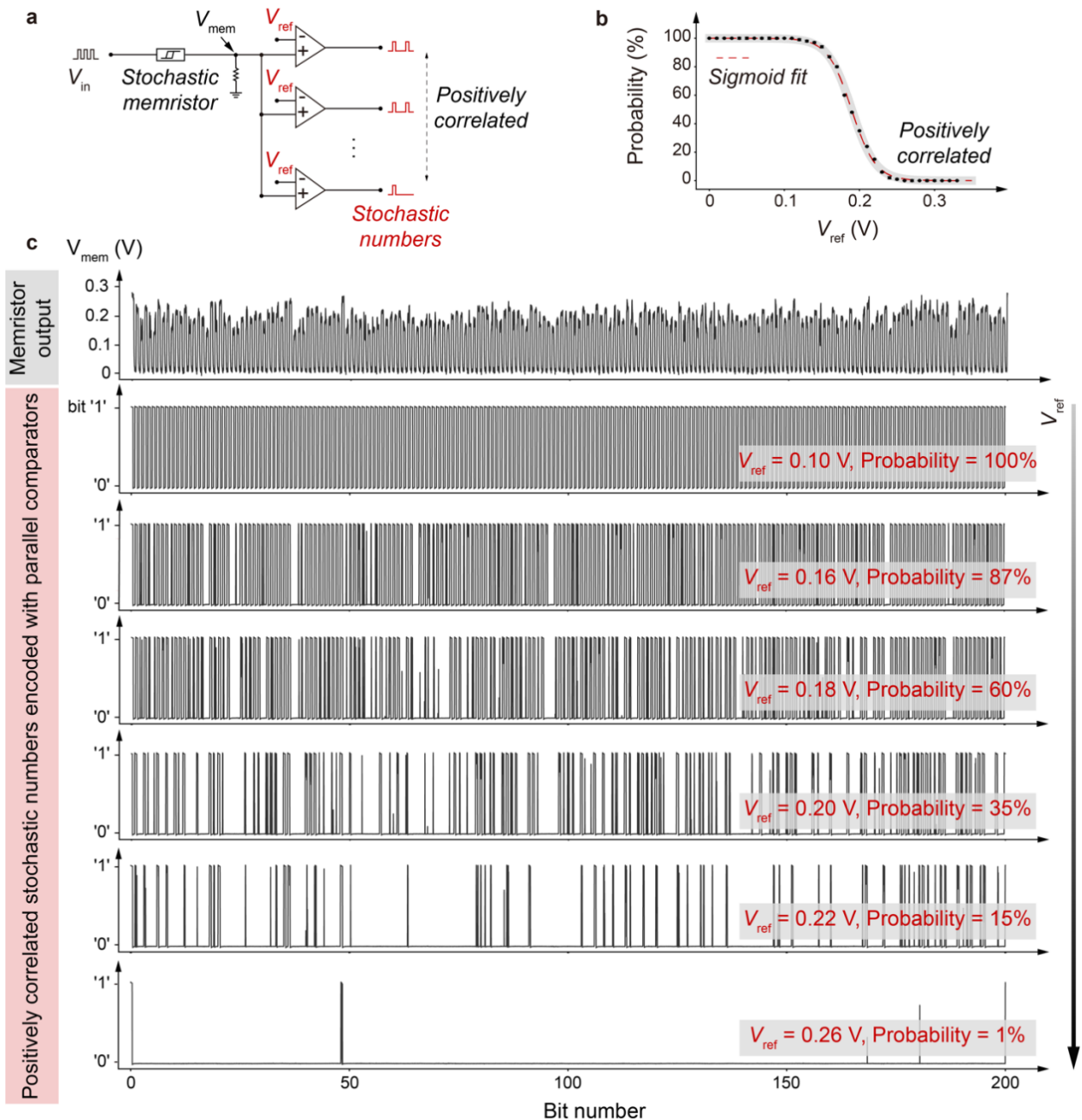

**Figure S6. Example showing the encoding of positive correlated stochastic numbers.** (a) Schematic SNE in positive correlation, replotted from Fig. 2a, to encode the positively correlated stochastic numbers. (b) Probability-$V_{ref}$ relation of the SNE in positive correlation, replotted from Fig. 2g, and (c) the corresponding experimental results. For a fixed memristor voltage output $V_{mem}$ (the first row), the SNE configured with two parallel comparators encodes positively correlated stochastic numbers. The probability of the encoded stochastic numbers, i.e. the probability of the '1' in the stochastic numbers, decreases as $V_{ref}$ increases, aligned well with the sigmoid fit in (b).

**a** *Example - when s is uncorrelated with a and b*

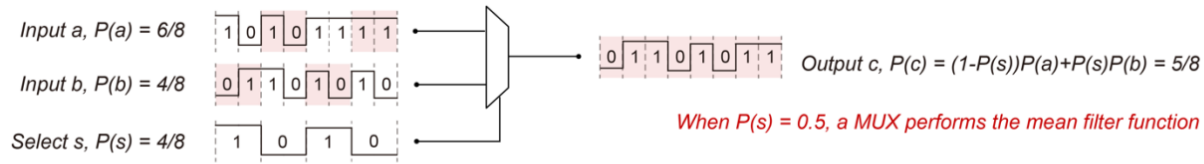

Input a, P(a) = 6/8

Input b, P(b) = 4/8

Select s, P(s) = 4/8

Output c, P(c) = (1-P(s))P(a)+P(s)P(b) = 5/8

*When P(s) = 0.5, a MUX performs the mean filter function*

**b** *Counter example - when s is correlated with a and/or b*

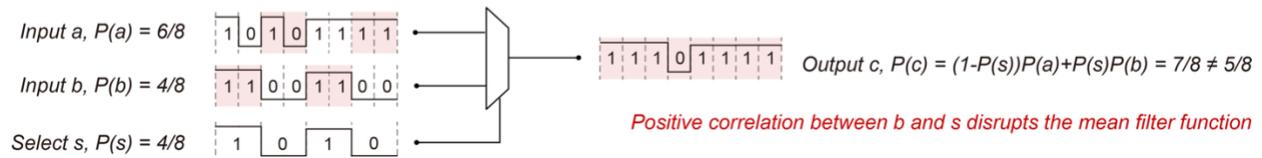

Input a, P(a) = 6/8

Input b, P(b) = 4/8

Select s, P(s) = 4/8

Output c, P(c) = (1-P(s))P(a)+P(s)P(b) = 7/8 ≠ 5/8

*Positive correlation between b and s disrupts the mean filter function*

**Figure S7. Stochastic MUX logic.** (a) An example showing the working principle of the stochastic MUX logic as a mean filter. Given that stochastic MUX logic follows $P(c) = \big(1 - P(s)\big)P(a) + P(s)P(b)$ as summarized in Table 1, a MUX can perform the mean filter function when $P(s) = 0.5$, i.e. $P(c) = 0.5(P(a) + P(b))$. Note that in this case the select $s$ should be uncorrelated with the inputs $a$ and $b$, and that the frequency of $s$ is half of that of the inputs to ensure that both the inputs participate in the logic operations. For example, as shown, $s$ selects bits from two input channels according to the bit level of $s$, where the bits as marked in blue are selected. As such, the MUX outputs the mean value (5/8) of inputs (6/8 and 4/8). (b) A counter example. A MUX no longer performs the mean filter function when the select $s$ is correlated with the inputs $a$ and/or $b$. As shown, a strong positive correlation between $b$ and $s$ disrupts the mean filter function, because $s$ completely accepts $b$ as part of the output $c$, instead of selecting bits from $b$ with a probability.

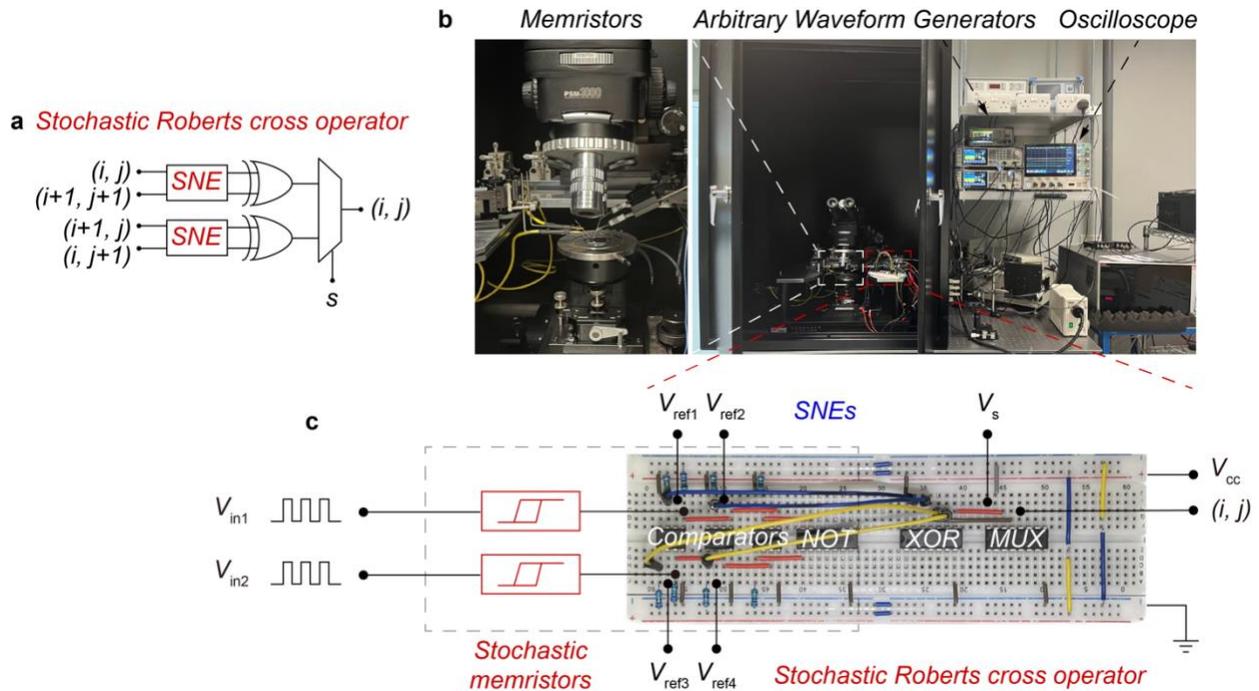

**Figure S8. The circuit design and experimental hardware realization of the stochastic Roberts cross operator.** (a) Circuit design (replotted from Fig. 4b). (b) Experimental setup. (c) Hardware realization. To build the stochastic Roberts cross operator, the memristors are tested on a probe station and connected to the logic gates and other electronic devices on a breadboard. The electronic components for the SNE realization include comparators, NOT gates, and resistors. The reference voltages of the comparators $V_{ref}$, the voltage supply $V_{cc}$, and the pulsed voltage signals $V_{in}$ are powered by the arbitrary waveform generators. The output of the stochastic Roberts cross operator $(i, j)$ is measured by the oscilloscope. Note that the frequency of the select $V_s$ of the MUX is half of that of $V_{in}$ to ensure that both the inputs participate in the logical operations.

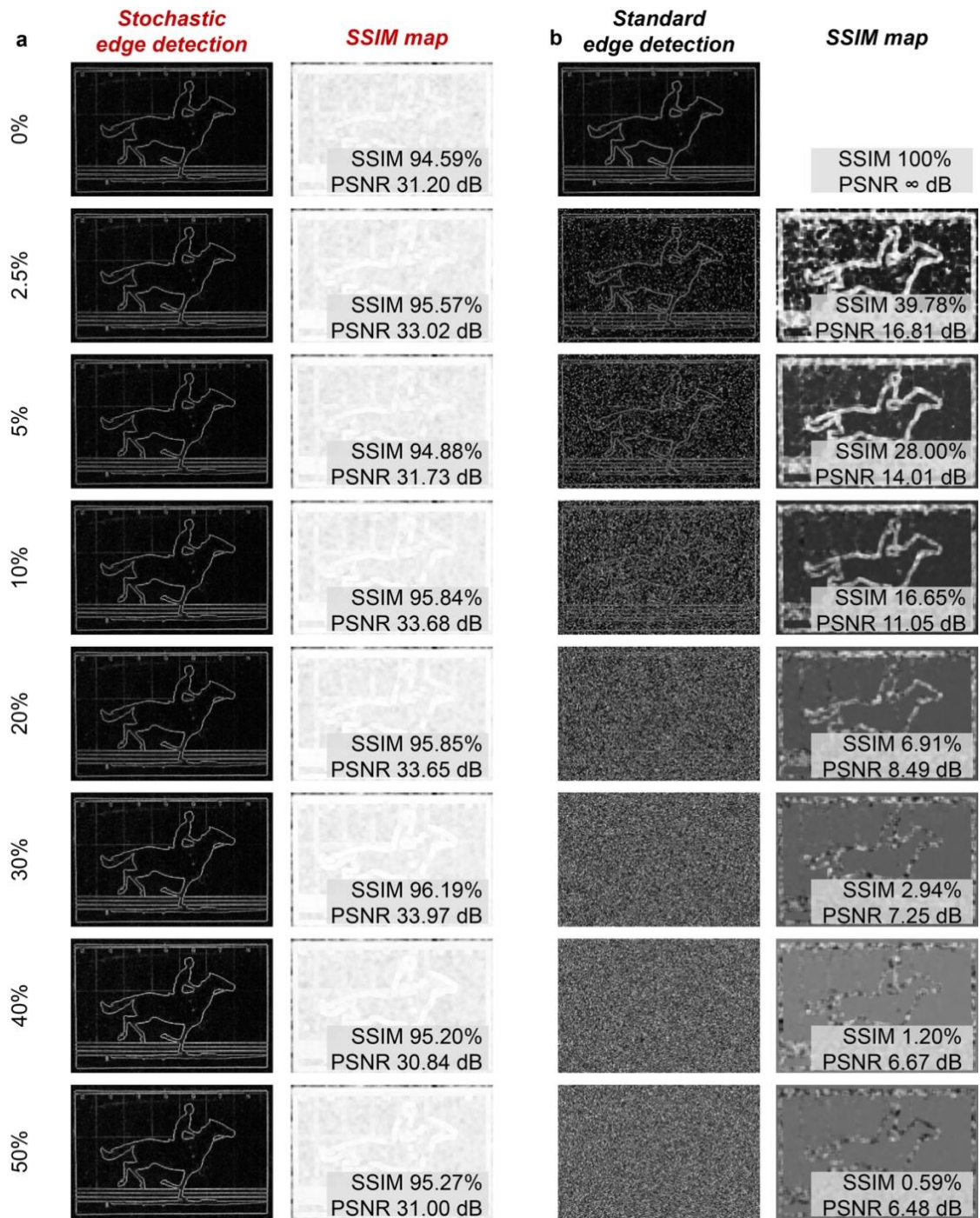

**Figure S9. Error-tolerance test.** (a) Stochastic and (b) standard edge detection results and the corresponding SSIM maps with bit-flip injection at a ratio of 0%, 2.5%, 5%, 10%, 20%, 30%, 40%, and 50%. For the stochastic edge detection, the high SSIM (>90%) and PSNR (>30 dB) prove that the bit-flip injection does not degrade the edge detection performance. In contrast, a low level of bit-flip injection significantly degrades the performance of the standard algorithmic edge detection.

## Supplementary References

1. L. Song, P. Liu, J. Pei, F. Bai, Y. Liu, S. Liu, Y. Wen, L. W. T. Ng, K. P. Pun, S. Gao, M. Q. H. Meng, T. Hasan, G. Hu, Spiking neurons with neural dynamics implemented using stochastic memristors. *Adv. Electron. Mater.* **2300564**, 1–9 (2023).
2. S. Dutta, G. Detorakis, A. Khanna, B. Grisafe, E. Neftci, S. Datta, Neural sampling machine with stochastic synapse allows brain-like learning and inference. *Nat. Commun.* **13**, 2571 (2022).